

\magnification=1200
\baselineskip=14pt

\def\cl{\centerline}

\ \

\bigskip\bigskip
\bigskip\bigskip
\bigskip\bigskip

\cl{\bf DETECTION OF SIGNATURE CONSISTENT WITH }
\cl{\bf COSMOLOGICAL TIME DILATION IN GAMMA-RAY BURSTS }

\bigskip\bigskip
\bigskip\bigskip

\cl{ J.P. Norris$^{(1)}$, R.J. Nemiroff$^{(1),(2)}$,
J.D. Scargle$^{(3)}$, C. Kouveliotou$^{(2),(4)}$, }
\cl{ G.J. Fishman$^{(4)}$, C. A. Meegan$^{(4)}$,
W.S. Paciesas$^{(4),(5)}$, and J.T. Bonnell$^{(1),(6)}$ }

\bigskip\bigskip
\bigskip\bigskip

\cl{\it $^{(1)}$ NASA Goddard Space Flight Center, Greenbelt, MD 20771 }
\cl{\it $^{(2)}$ Universities Space Research Association }
\cl{\it $^{(3)}$ NASA Ames Research Center, Moffet Field, CA 94035 }
\cl{\it $^{(4)}$ NASA Marshall Space Flight Center, Huntsville, AL 35812 }
\cl{\it $^{(5)}$ University of Alabama, Huntsville, AL 35899 }
\cl{\it $^{(6)}$ Computer Sciences Corporation }

\bigskip\bigskip
\bigskip\bigskip
\bigskip\bigskip
\bigskip\bigskip
\bigskip\bigskip
\bigskip\bigskip

\cl{ In press: ApJ, 1 April 1994 }

\vfill\eject
\baselineskip=18pt

\ \
\bigskip\bigskip

\cl{\bf ABSTRACT }

\bigskip

If gamma-ray bursters are at cosmological distances - as suggested by their
isotropic distribution on the sky and by their number-intensity relation -
then the burst profiles will be stretched in time, by an amount
proportional to the redshift, $1 + z$.  We have tested data from the
Compton Gamma Ray Observatory's Burst and Transient Source Experiment
(BATSE) for such time dilation.  Out of 590 bursts observed by BATSE, 131
bursts were analyzed; bursts with durations shorter than 1.5 s were
excluded.  We used three tests to compare the timescales of bright and dim
bursts, the latter, on average, being more distant than the former.  Our
measures of timescale are constructed to avoid selection effects arising
from intensity differences by rescaling all bursts to fiducial levels of
peak intensity and noise bias.  (1) We found that the total rescaled count
above background for the dim burst ensemble is approximately twice that for
the brightest bursts - translating into longer durations for the dim
bursts. (2) Wavelet-transform decompositions of the burst profiles
confirmed that this dilation operates over a broad range of timescales.
(3) Structure on the shortest timescales was examined using a procedure
which aligns the highest peaks of profiles from which the noise has been
optimally removed using a wavelet thresholding technique.  In all three
tests, the dim bursts are stretched by a factor of $\sim$ 2 relative to the
bright ones, over seven octaves of timescale.  We calibrated the
measurements by dilating synthetic bursts that approximate the temporal
characteristics of bright BATSE bursts.  Results are consistent with bursts
at BATSE's peak-flux completeness limit being at cosmological distances
corresponding to $z \sim 1$, and thus with independent cosmological
interpretations of the BATSE number-intensity relation.  Alternative
explanations of our results, arising from the nature of physical processes
in bursts, are still possible.

\bigskip

\noindent {\it Subject headings:} cosmology:  theory - gamma rays:  bursts

\vfill\eject

\cl{\bf 1. INTRODUCTION }

\bigskip

The long-standing mystery of the nature and distances to gamma-ray burst
sources might be addressable in the near future by simultaneous detection
of a single burst counterpart in another waveband.  Such a detection could
serve as the ``Rosetta Stone" of bursts, especially if the bursters are a
galactic halo and/or disk population.  It is less likely that counterparts
at cosmological distances could be studied in quiescence, especially since
some critical determinants might not be observable:  the best determined
burst error positions for the brightest bursts do not appear to be
associated with bright galaxies (e.g., luminosity comparable to M31) or
with AGNs (Fenimore et al. 1993).  Instead, a good proof of cosmological
origin would be multiple gravitationally lensed (temporally separated)
images of a burst.  A few such cases might be seen over the lifetime of the
Compton Observatory, depending on the depth of a cosmological distribution
(Nemiroff et al. 1993b).  The cosmological origin of bursts would also be
indicated by detection of galactic longitude/latitude-dependent absorption
in the soft X-ray band.  It is quite unlikely, however, that any
measurement of the distribution of source distances will be made in the
near future - save by analyzing statistical attributes of the gamma-ray
bursts themselves.  In fact, Compton's Burst and Transient Source
Experiment (BATSE) may be sampling deep enough to see the effects of non-
Euclidean space.  The practically isotropic celestial distribution and the
$V/V_{\rm max}$ distribution for gamma-ray bursts detected and localized
(Meegan et al. 1992) by BATSE are most naturally interpreted in terms of
either a cosmological (Mao \& Paczynski 1992; Wickramasinghe et al. 1993)
or heliospheric distribution of sources.

If the bursters are at cosmological distances, then the time
profiles of more distant sources will be dilated relative to those of
nearer sources (Paczynski 1992; Piran 1992; see also McCrea 1972). Since
burst durations range over more than four orders of magnitude, with modes
near 25 s and less than 1 s (Kouveliotou et al. 1993), dilation would be
detectable only in a statistical sense. Most of this range in duration
arises from intrinsic variations in the rest frames of the sources; only a
factor of $\sim$ 2 would be expected from cosmological expansion.  A positive
finding would not prove outright the cosmological hypothesis, but would
increase the burden on any Galactic scenario, requiring additional ad hoc
theories to explain an anticorrelation of intensity with duration.  In
addition, the spectral energy distribution of a burst will be redshifted -
by the same factor as the temporal profile is dilated.  The redshift effect
should be detectable at energies below where the burst spectra begin to
exhibit curvature.  At higher energies, in the quasi-power-law regime,
detection is problematic.

We have applied several methods to test for time dilation in bursts
detected by BATSE, and report here the results of three tests which are
robust in that the estimated timescales are nearly independent of the
apparent brightness of the burst.  The tests also address the problem of
the wide distribution of temporal structure observed in bursts.  We
describe the preparation of burst profiles to achieve a uniform noise bias,
the application and results of each test, and comparisons with simulations.
We then discuss necessary refinements and the consistency of our results
with a cosmological interpretation of the number-intensity relation
(Wickramasinghe et al. 1993) as determined by BATSE.  Our measurements,
which do not yet take into account redshift effects on temporal profiles,
yield a consistent picture of a relative time dilation factor of order 2
between brightest and dimmest bursts, and would thus place the more distant
bursts at redshifts of order unity.  The model-dependent effects of
redshift will be addressed in a future paper.

\bigskip

\cl{\bf 2. ANALYSIS PROCEDURES AND RESULTS }

\bigskip

We utilized burst time profiles recorded with 64 ms resolution from BATSE's
Large Area Detectors (LADs).  In the initial phase of the mission, burst
data acquisition began about 2 s prior to burst trigger (that instant when
the experiment senses that a significant rise in count rate above
background has occurred) and continued to $\sim$ 240 s after (see Fishman et
al.
1989, Kouveliotou et al. 1993, for descriptions of instrument and data
types); since 1993 February, about 1000 s of posttrigger data have been
recorded for each burst.  The 64 ms data are summed over the subset of
detectors which trigger on the burst.  We concatenated prior to the 64 ms
data four samples from a data type with lower time resolution (1.024 s),
which is continuously available, summing over the same triggered detectors.
(For the peak alignment test described in \S 2.3 we concatentated 16
such samples.)  This measure includes structure that would otherwise be
lost in some dim bursts which commence and rise slowly before triggering
the experiment's higher time resolution modes.  Each of the 1.024 s
pretrigger samples was divided into sixteen 64 ms samples, with the counts
randomly distributed in time and rebinned to the finer resolution.
Utilization of additional 1.024 s data prior to trigger would sometimes
result in loss of information in long bursts since, to minimize the effects
of background curvature (see below) and to realize an interval with 2n
points for the wavelet spectrum test, we truncated the concatenated time
profiles at 1024 samples (65 s).  Bursts shorter than 1.5 s were excluded
from the analysis for two reasons.  First, the BATSE sensitivity is lower
for very short, dim bursts.  Also, a bimodal burst duration distribution
(see Mazets et al. 1981; Norris et al. 1984) is now confirmed by BATSE, and
it is possible that bursts shorter than about 1 s constitute a distinct
phenomenon or subpopulation (Kouveliotou et al. 1993).  Out of a total
sample of 590 bursts observed by BATSE, 131 were analyzed; 24 with
durations shorter than 1.5 s were excluded.  With the number of presently
available bursts it would be difficult to determine a priori at what
duration to truncate a sample of dim or bright bursts drawn from the short
variety.  Gaps occurring near background level were tolerated and filled
with Poisson noise with the mean rate computed from intervals adjacent to
the gap.  The effects of dead time in our analysis are negligible for the
burst sample used.

Some measure of burst luminosity must be assumed for normalization
and interpretative purposes.  For this we used the peak of the counting
rate at 256 ms resolution summed over the four LAD energy channels
(approximately:  25 - 50 keV, 50 - 100 keV, 100 - 300 keV, and $>$ 300 keV).
A background interval was defined using 64 ms data from an interval
(typically 96 s in duration) after cessation of the burst.  The background
was fitted with a first-order polynomial and subtracted from the burst
rates.  If the extrapolated fit did not also closely fit that portion of
the pre- trigger data which was (apparently) not part of the burst, the
background interval was redefined and the procedure iterated.  Since bursts
often contain temporal structures with heterogeneous widths and shapes, a
method for estimating the peak intensity in a way that is unaffected by a
rescaling of the time axis is desirable.  Donoho \& Johnstone (1993)
describe such an algorithm: (1) perform a wavelet transform on the time
series; (2) determine a threshold (that is allowed to be different at each
scale) using a risk minimization procedure; (3) delete wavelets with
amplitudes that are below the threshold and diminish the others by the
threshold value; and (4) reconstitute the time series by an inverse wavelet
transform.  At sufficiently high signal-to-noise (S/N) levels, the
algorithm tends to preserve the amplitude of both broad and narrow features
while significantly reducing noise fluctuations; however, intensity loss
that depends on feature width is incurred if the S/N is too low.  We
adapted this noise reduction method for the time dilation tests, tailoring
the thresholding scheme to minimize the intensity-loss problem which arises
for dim bursts.  We used the Haar basis (\S 2.2) to implement wavelet
thresholding.  We chose to use the data smoothed to 256 ms resolution for
estimating the peak intensities because on shorter timescales preservation
of structure in a manner independent of timescale is problematic for the
dimmest bursts.  Using simulations we demonstrated that for all bursts
analyzed, the resulting peak intensity loss is 2 \% or less across the
observed range of structure widths.  We selected bursts that fall into
three brightness groups, with the extremes differing by more than two
orders of magnitude in peak intensity (above background), as shown in Table
1.

In preparation for all tests the burst peak intensities, backgrounds, and
noise biases were rendered uniform, with the aim of virtually eliminating
brightness selection effects.  After subtraction of the fitted background
(Bb), the remaining ``signal" profile was diminished to a canonical peak
intensity (that of the dimmest burst used, 1400 counts s-1), and a
canonical flat background (Bc) was added (the average background for the
dimmest bursts, 5600 counts s-1). Over the fixed time interval analyzed (65
s), the actual backgrounds are relatively free of curvature and fairly
constant among dimmer bursts.  Backgrounds among the bright bursts vary by
more than a factor of 2, but this additional variance is negligible after
the signal level has been diminished.  The reduced signal plus constant
background (Ri) was then fractionally Poisson- distributed to compensate
for reducing the original time series.  The preparation procedure is thus
expressed:
 $$                R_i  =  f (C_i  -  B_b) + B_c + V_f
 \eqno(1)$$
where $f$ is the signal reduction factor and Ci is the original sample value
for interval $i$.  The required fractional variance, $V_f$ , is obtained from
 $$             d_i  =  R_i (1 - f) + f(B_c - f B_b)
 \eqno(2)$$

 $$             V_f  =  P(d_i) - d_i ,
 \eqno(3)$$
where $P(d_i)$ is a Poisson variable (deviate) whose mean value is $d_i$.
This procedure would result in an exact realization of a
Poisson-distributed time series - with signal and background levels
rendered constant - except that we must set $B_b = B_c$ to avoid negative
deviates.  Because the simulated Poisson noise arising from the canonical
background is white, whereas the actual subtracted background sometimes
contains small but nonnegligible lower frequency components, for the
wavelet decomposition analysis (\S 2.2) small corrections of order 2 \% or
less were added per brightness group.  The corrections were generated by
comparing the wavelet decompositions of actual background noise and of
simulated noise.  The actual noise intervals were taken from the 64 ms data
records of dim bursts which appear to have subsided soon enough to afford a
flat 65 s background interval.  The judgment to utilize a particular
background interval for this purpose was made by fitting and subtracting a
first-order polynomial and then examining the trend of residuals expressed
in standard deviations.

Concomitant with time dilation is spectral energy redistribution due to
redshift, which would also affect temporal measurements:  Because pulses
tend to be narrower at higher energy, redshifted bursts would have these
narrower structures shifted to lower energies.  We wish to minimize this
possible effect for the present analysis.  At lower energies, the disparity
between pulse widths as a function of energy is not as large, and therefore
redshift is less of a complicating factor.  From fits to 45 isolated pulses
in BATSE bursts we find the average pulse-width ratios, channels 2:1 and
3:2, to be roughly 0.85 in both cases.  This is not enough of a departure
from unity to mask a dilation factor of order 2; that is, temporal
structure in the rest frame in the 100 - 200 keV band of bursts from
distant sources redshifted to 50 - 100 keV in the observed frame would be
only $\sim$ 15 \% narrower than structure in bursts from nearby sources in
the 50 - 100 keV band.  Therefore, to perform the time dilation tests
reported here, we used data summed over channels 1 and 2 (25 - 100 keV).
(Pulse widths in channel 4 are often much narrower.  If dim bursts are
redshifted by approximately a factor of 2, narrower structure is shifted
into channel 3. However, we utilized all four channels in estimating the
peak intensity in order to minimize uncertainties.)

\bigskip

\noindent
2.1  Total Normalized Counts Test

The total count, or fluence, in each prepared time profile was measured by
subtracting the constant canonical background (variance arising from
background is retained) and summing over the 65 s interval.  Sample errors
for this measure were then determined for each brightness group. Included
in Table 1 are the number of bursts (N), the average total counts (S) above
background summed over channels one and two, and the sample errors.  All
bursts are normalized to the same canonical peak intensity (flux) of 1400
counts s-1 and therefore S  (fluence) can be given meaning in terms of an
average ``equivalent duration" - average ratio of fluence to flux.  The
average equivalent durations, teq, for the two dim groups, 6.6 and 6.0 s,
are about twice as long as that for the bright group, 3.2 s. The difference
is significant at about the 3.5 s level.  Two-s fluctuations in S translate
into a range in the ratio of equivalent durations of $\sim$ 1.5 - 3.

This linear average of the total counts per brightness group
indicates the presence of a significant integrated effect, prompting the
question:  Is such a stretching of temporal features observed on each
timescale present in gamma-ray bursts?  Note that the same stretch factor
must be observed on all timescales for the signature to be consistent with
cosmological time dilation.  The remaining two tests address the scaling
question on timescales from 256 ms to 65 s, making more complete use of the
information present in the burst profiles.

\bigskip

\noindent
2.2.  Wavelet Spectrum Test

The second test employs Haar wavelet transforms (see Daubechies 1992) to
measure the amount of structure in bursts on a range of timescales.  Since
wavelet transforms are indexed in position and timescale, they may be
superior to, for example, Fourier transforms for characterizing the
spectrum of a time series which contains a variety of asynchronous,
heterogeneous structures - often the case in gamma-ray bursts.  The Haar
wavelet is a difference operator.  Among orthonormal wavelet bases, the
Haar basis affords the most compact support, differencing only two adjacent
intensity samples on a given scale, thereby preserving discontinuous
features with the highest fidelity (however, it is not necessarily the
optimum basis for characterizing the structure in gamma-ray burst
profiles).  A complete Haar wavelet transform is a decomposition of the
profile into localized intensity differences, on a range of timescales,
plus a constant component, the mean value.  On timescale m and at position
n, the wavelet transform of the function f(t) can be expressed (Daubechies
1992)
 $$   T_{m,n}(f)  =  a_0^{-m/2} \int dt f(t) \psi(a_0^{-m_t} - nb_0) ,
 \eqno(4)$$
with constants $a_0  > 1$ and $b_0  > 0$.  For our analysis $y$ is the Haar
wavelet:
 $$ \eqalign{ y(t)&  =   ~~1  ~~~~~ 0 \le t < 1/2 \cr
                  &  =   -1  ~~~~~1/2 \le t < 1 \cr
                  &  =   ~~0  ~~~~~ {\rm otherwise.} \cr }
 \eqno(5)$$
For discretely spaced data, the time series to be differenced on each
successive scale can be generated by summing pairs of intensity samples
(``compressing" by a factor of 2).  In this analysis the temporal range
from 128 ms to 65 s is spanned in dyadic steps.

Our measure of temporal structure, or activity, on a given scale is the
average of absolute values of differences between pairs of intensity
values.  We call this average over the position index of the wavelet
transform a ``wavelet amplitude spectrum."  In analogy with the Fourier
spectrum, phase (positional) information is lost in the summation process.
Figures 1a and 1b illustrate the average wavelet spectra for the three
brightness groups.  The practically Poisson noise is flat, at a value of
15.1 units.  We applied a small correction in computation of the noise
bias, resulting in negligible differences in the curves on timescales less
than 2 s (see Norris et al. 1993b).  The average wavelet spectrum for the
bright bursts (dashed line) lies below that for the other two groups
(dotted lines) on timescales from $\sim$ 4 to 65 s.  Empirical sample
errors for each group are shown as envelopes (solid lines).  On timescales
longer than about 2 s the error envelopes are mostly determined by the
ranges of durations and profile shapes within the three brightness groups.
On shorter timescales noise fluctuations begin to dominate the errors; the
reason for relatively smaller errors is that only a few significant short
timescale features are present in a burst of average duration, whereas many
more noise fluctuations are averaged to produce the 65 s wavelet spectrum.
The signal on short timescales is investigated using the third test (\S
2.3).

To calibrate the effect seen in Figure 1 we performed simulations of
gamma-ray bursts.  We fitted 220 pulses in 30 bursts.  The form of our
pulse model is (see Norris et al. 1993a):
 $$     F(t)  =  A  \ {\rm exp} ( - ( |t - t_{max} | / \sigma_{r,d} )^\nu ) ,
 \eqno(6)$$
where $t_{max}$  is the time of the pulse's maximum intensity, $A$;
$\sigma_r$ and $\sigma_d$ are the rise ($t < t_{max}$) and decay time
constants ($t > t_{max}$), respectively; and $n$ is the pulse ``peakedness"
(a lower value for $n$ results in a more peaked pulse).  Combined, the
parameters $\sigma_r$, $\sigma_d$, and $n$ afford sufficient flexibility to
represent virtually all pulse structure shapes in bursts. Most of the
fitted pulses overlap substantially, nevertheless their sum faithfully
represents the total profile on timescales longer than about 1 s
(significant, shorter timescale components comprise the fit residuals in
some cases).  The simulations approximate distributions of several
fundamental properties of burst profiles:  duration, number of pulses per
burst, pulse rise and decay times, and pulse peakedness. The tendency for
pulses to cluster was not simulated for results discussed here.  Bright
bursts were simulated from the parent population of fitted pulses,
preserving all fitted parameters for each burst except for the positions of
pulses, which were randomized for each realization.  Dim bursts were
simulated in the same manner with these modifications:  rise and decay
times and burst durations were dilated by a constant factor.  Each
simulated profile with canonical background level added was Poisson
distributed.

Figure 2 illustrates the results of applying the wavelet spectrum test to
these simulations.  The average spectra are for 100 dilated and 100
nondilated burst simulations where the relative dilation factor is 2.25.
The simulated curves in Figure 2 agree fairly well with those in Figure 1,
given the extents of the error envelopes.  A higher curve indicates more
structure, which may be manifested - under the circumstances of our
intensity normalization - by a longer average duration. Essentially, the
(fixed length) 65 s interval is more filled up per timescale for the
dilated bursts of Figure 2 and for the dimmer bursts of Figure 1.  This may
seem counterintuitive; dilation might have been expected to move a given
curve to the right rather than up.  For simplicity, consider only processes
composed of Haar wavelets of constant amplitude, independent of their
timescale, $\tau_H$, and which follow number distributions of the form
 $$  N_H(\tau_H)  \propto  \tau_H^{\alpha}
 \eqno(7)$$

For processes with $\alpha < 0$, a dilation transformation with $\kappa  >
1$, $N_H(\tau_H) \rightarrow N_H'(\kappa \tau_H)$, results in an increased
number of wavelets on the dilated timescale by a factor
$\tau_H^{\alpha}/(\kappa \tau)^{\alpha} = \kappa^{-\alpha}$.  Conversely,
dilation for processes with $\alpha > 0$ results in fewer wavelets on the
longer timescale.  In reality there must be a dependence of the number
distribution on amplitude as well as timescale, and a simple power law as
in equation (7) probably does not apply.  For the gamma-ray burst process,
dilation apparently moves {\it more} structure to the new (dilated)
timescale, and this appears in the wavelet spectrum as an increase in
summed amplitude.  This demonstrates the necessity of simulating the
distribution of structure in bursts with sufficiently high fidelity.

\bigskip

\noindent
2.3.  Peak Alignment Test

On timescales shorter than 2 - 4 s the wavelet spectrum test is virtually
defeated by noise.  To realize a test relatively free of noise bias for the
shorter timescales, we introduce a refinement of the ``alignment of highest
peaks" technique advanced by Mitrofanov et al. (1993).  After having
prepared the time profiles as described in \S 2, we applied the same
wavelet-thresholding technique used in the procedure to estimate peak
intensity, thereby eliminating noise fluctuations.  We then shifted the
thresholded profiles in time, bringing the highest peaks of all profiles
into common alignment.  The peak-aligned profiles were then averaged for
each brightness group.  Without thresholding, the region near the average
aligned peak would be contaminated by noise (since all bursts are reduced
to the canonical dimmest peak intensity), and the test would be
compromised. Since we are interested in preserving information near the
highest peak in a symmetric manner, sixteen 1.024 s samples are
concatenated prior to the burst trigger data for this test.  We used burst
profiles with 128 ms resolution in an attempt to observe the stretching
effect on timescales as short as the S/N allowed.

Figure 3 shows the results of this test for the region within 8 s of the
average peak.  The structure near the peak is necessarily the most
significant within a given burst, but the widths and shapes of the
contributing structures in the vicinity of the average peak vary greatly.
Thus, like the wavelet spectrum test, this peak alignment test examines a
range of timescales.  Unlike the former test, it incurs little noise bias
because thresholding has eliminated the majority of insignificant
fluctuations and the most significant structure has been selected by the
alignment process.  Using simulated sets of bursts, we estimated the
intensity loss arising from thresholding to be 10 \% or less for structures
with full width half-maxima of 0.5 - 1.0 s (the mode for highest peaks) for
the sum of channels one and two.  The average profile ratio of sets with
and without Poissonization and wavelet thresholding procedures provided an
estimate for the intensity loss correction.  These corrections were applied
to the curves in Figure 3.  The peak of the average profile for the bright
group is significantly lower than those of the dim groups.  This is not a
noise effect, but must result from the bright bursts having lower peak
count rates in channels 1 and 2 than do the dimmer bursts (recall that the
profiles are normalized to the peak intensity of all four channels summed).
Spectral energy redistribution arising from redshift may be the explanation
for this effect.  There is a clear trend on timescales $\tau_w > 0.5$ s:
across five octaves in width the average profile for the bright bursts is
approximately one-half as wide as the average profiles for the dim and
dimmest bursts.  If we normalize the peak of the average profile for the
bright bursts to unity, the differences in width of the average profiles
are still significant for $\tau_w > 0.5$ s.  If the deficit for bright
bursts is due to spectral redshift, then a self-consistent treatment would
also require propagation of the narrower temporal structures in higher
energy bands to lower energy bands, as discussed in \S 2.

We estimated the uncertainties of the peak alignment test by performing one
100 sets of simulations in the manner described above, each set consisting
of 50 dilated and 50 nondilated profiles, again with a constant dilation
factor of 2.25.  Figure 4 illustrates the application of the peak alignment
test to these simulations.  The envelopes (solid lines) contain 68\% of the
profiles per group.  From visual inspection, we conclude that the
differences in the average profiles of dim and bright bursts in Figure 3
are significant at about the 2 to 4 s level per octave in width (although,
each octave in width does not represent an independent quantity as in the
wavelet spectrum test).  As noted previously, our simulations do not
reproduce all of the narrow structure observed in actual bursts, hence in
Figure 4 the profile uncertainties within $\sim$ 0.5 s of the average peak are
not well modeled.

\bigskip

\cl{\bf 3. DISCUSSION }

\bigskip

We selected bright and dim gamma-ray bursts detected by BATSE according to
their peak intensities.  We performed a procedure to render uniform the
burst signals, backgrounds, and noise biases, thus virtually eliminating
any brightness selection effects.  An ensemble measure of the fluence to
flux ratio is twice as large for dim bursts as for bright ones, suggesting
that to first order, the dimmer bursts have durations which are twice as
long.  Combined, a wavelet spectrum test and a peak alignment test
calibrated with burst simulations confirm that a ``stretch" ratio of $\sim$
2 is manifest over seven octaves in width of temporal structure, from 0.5
to 65 s.  The efficacy of the latter two tests is that they segregate
(wavelet spectrum) or order (peak alignment) the information in a time
profile by width of temporal structure, and thus tend to address the
dispersion which arises from the wide variety of time profiles.  The
capability to resolve profiles into components actually takes advantage of
the intrinsic dispersion:  the stretching effect must be present on all
timescales to be consistent with cosmological time dilation.  The peak
alignment test affords a stretch measure on shorter timescales than does
the wavelet spectrum, essentially because the most significant narrow
features of bursts are selected and averaged.  Wide ($>$ 8 s),
low-intensity structures are not as well estimated by the peak alignment
test because our noise thresholder has been optimized for intense features.
 The wavelet spectrum test is more accurate for measurements on long
timescales since broader burst structures, containing more counts, dominate
over noise.

We note that Mitrofanov et al. (1993) designed a similar peak alignment
procedure to search for effects like time dilation and found a negative
result when applied to bursts detected by the APEX experiment on PHOBOS 2.
We believe there is not necessarily an inconsistency with our result since
their dynamic range in peak intensity was approximately an order of
magnitude less than BATSE's, suggesting that APEX did not sample
sufficiently distant sources to observe the effect. Lestrade et al. (1992)
found a positive effect in a subsample of short bursts detected by PHEBUS
(again with a much smaller dynamic range in peak intensity than BATSE's),
but given the clear bimodality in the burst duration distribution with a
minimum near 1.5 s (Kouveliotou et al. 1993) - equivalent to the duration
cutoff of their sample, it is difficult to interpret their result.

Some effects which we do not yet take into account would tend to increase
the observed differential signal.  As explained in \S 2, redshift of the
energy spectrum is a competing effect since at higher energy, pulses tend
to be narrower.  A temporal analysis taking into account such a ``K-
correction" is somewhat more complex, and dependent upon an assumed
cosmology.  From energy-dependent pulse-shape fits, we estimate correction
for redshift would tend to increase the apparent time dilation effect by 10
\% - 20 \%.  However, spectral redistribution arising from redshift would
tend to decrease the effect by a comparable amount.  A precise background
model would allow the use of time series longer than 65 s, and the analysis
of the complete event in practically all cases.  This would tend to include
preferentially more structure for dim bursts since they are, on average,
longer.  Also, since we have fitted backgrounds (apparently) after burst
cessation and since bursts tend to be asymmetric on all timescales
(Nemiroff et al. 1993a), such that burst ``envelopes" decay more slowly than
they rise, for dim bursts our background subtraction will tend to remove
some signal if the event decays slowly enough.  Better background
estimation will alter our results only on the longest timescales.  We will
address these problems in a future paper.

Finally, the effects of an intrinsic luminosity distribution must be
modeled if more precise comparisons are to be had with interpretations
(Fenimore et al. 1993; Wickramasinghe et al. 1993) of burst
number-intensity relations determined by BATSE and other instruments.
However, we can make a coarse comparison of our result with such analyses
at this stage.  For the first 260 bursts detected by BATSE, the average
values of Cmax/Cmin for the bright, dim, and dimmest groups are 34.5, 2.10,
and 1.75, respectively.  The Cmax/Cmin parameter is about a factor of 2
larger than the Cpeak/Ccom measure used in Wickramasinghe et al. (1993).
{}From this we see that our dimmest group is near the 99\% completeness limit
for BATSE ($C_{\rm peak}/C_{\rm com} = 1$), which, as determined from
Figure 2 of Wickramasinghe et al., corresponds almost exactly with a
redshift of one. Therefore, we consider our result, $z \sim 1$, to be in
good agreement with independent fits to the BATSE number-intensity
relation.

Given the state of knowledge of burst mechanisms, the anticorrelation of
brightness with duration admits alternative explanations.  For example, if
the mechanism is such that total energy per burst is approximately
constant, then release of energy at a slower rate, corresponding to a lower
luminosity, would result in longer bursts.  However, this exact
proportionality is not what we observe.  Also, if the bursters are
cosmological, it is entirely possible that evolutionary effects complicate
the picture; for example, the duration and peak intensity distributions
could evolve with time.

\vfill\eject

\def\tabletoprule{\noalign{\hrule\smallskip}}
\def\tablerule{\noalign{\smallskip\hrule\smallskip}}
\def\tablebottomrule{\noalign{\smallskip\hrule}}

\cl{ Table 1 }
\cl{ Total Normalized Counts Test }

\bigskip\bigskip

\halign{#\hfil& \quad\hfil#& \quad\hfil#& \quad\hfil#& \quad#\hfil&
                \quad#\hfil& \quad#\hfil& \quad\hfil#\cr
\tabletoprule
Brightness&  Peak~~&& N~~~& $\tau_{eq}$& $~\Sigma^b$& +1 $\sigma$&
-1 $\sigma$\cr
 ~~Group&    Intensity$^a$&&   (bursts)&  (s)\cr
&            cts s$^{-1}$~\cr
&         Min~&    Max~\cr
\tablerule
Bright&   18,000&  250,000&   41~~~&      3.2&     4490&    1050&    420\cr
Dim&       2,400&    4,500&   44~~~&      6.0&     8340&    1030&   1400\cr
Dimmest&   1,400&    2,400&   46~~~&      6.6&     9300&    1870&    1240\cr
\tablebottomrule}

\bigskip\bigskip

\noindent
$^a$ Peak intensity summed over LAD channels 1 - 4.

\noindent
$^b$ Average total counts summed over channels 1 and 2.

\vfill\eject

\cl{\bf REFERENCES }
{
\parindent=0pt
\hangindent=20pt
\baselineskip=10pt
\parskip=4pt

\bigskip

Daubechies, I.  1992, Ten Lectures on Wavelets (Philadelphia:  Capital City
Press)

Donoho, D., \& Johnstone, I. 1993, preprint

Fenimore, E.E., et al.  1993, Nature, accepted

\hangindent=20pt
Fishman, G.J., et al.  1989, in Proc. Gamma Ray Observatory Science
Workshop, ed W.N. Johnson (Greenbelt, MD: NASA/GSFC), p 2-39

\hangindent=20pt
Kouveliotou, C., Meegan, C.A., Fishman, G.J., Bhat, N.P., Briggs, M.S.,
Koshut, T.M., ApJ, in press

Paciesas, W.S., \& Pendleton, G.N.  1993, ApJ, 413, L101

Lestrade, J.P., et al.  1992, A\&AS, 97, 79

Mao, S., \& Paczynski, B.  1992, ApJ, 388, L45

Mazets, E.P., et al.  1981, Ap\&SS, 80, 3

\hangindent=20pt
McCrea, W.H. 1972, in External Galaxies and Quasi Stellar Objects, ed. D.S.
Evans, IAU, p 283

Meegan, C.A., Fishman, G.J., Wilson, R.B., Paciesas, W.S., Pendleton, G.N,
Horack, J.M., 1992, Nature, 355, 143

Brock, M.N., \& Kouveliotou, C.  1992, Nature, 355, 143

\hangindent=20pt
Mitrofanov, I., et al.  1993, in Proc. Compton Symp., ed. M. Friedlander,
N. Gehrels, \& D. Macomb (New York:  AIP), in press

\hangindent=20pt
Nemiroff, R.J., Norris, J.P., Kouveliotou, C., Fishman, G.J., Meegan, C.A.,
\& Paciesas, W.S. 1993a, ApJ, in press

Nemiroff, R.J., et al.  1993b, ApJ, 414, 36

Norris, J.P., Cline, T.L., Desai, U.D., \& Teegarden, B.J.  1984, Nature,
434, 308

\hangindent=20pt
Norris, J.P., Davis, S.P., Kouveliotou, C., Fishman, G.J., Meegan, C.A.,
Wilson, R.B., \& Paciesas, W.S.  1993a, in Proc. Compton Symp., ed. M.
Friedlander, N. Gehrels, \& D. Macomb (New York:  AIP), in press

\hangindent=20pt
Norris, J.P., Nemiroff, R.J., Kouveliotou, C., Fishman, G.J., Meegan, C.A.,
Wilson, R.B. \& Paciesas, W.S.  1993b, in Proc. Compton Symp., ed. M.
Friedlander, N. Gehrels, \& D. Macomb (New York:  AIP), in press

Paczynski, B.  1992, Nature, 355, 521

Piran, T.  1992, ApJ, 389, L45

\hangindent=20pt
Wickramasinghe, W.A.D.T., Nemiroff, R.J., Norris, J.P., Kouveliotou, C.,
Fishman, G.J., Meegan, C.A., Wilson, R.B., \& Paciesas, W.S. 1993, ApJ, 411,
L55

}

\vfill\eject

\cl{\bf FIGURE CAPTIONS }
\hangindent=0pt
\baselineskip=18pt

\bigskip

\noindent {\bf Figure 1:} Average wavelet amplitude spectra of BATSE bursts
for three brightness groups, described in Table 1:  (a) dim and bright, (b)
dimmest and bright groups (dotted and dashed respectively).  Solid lines
indicate empirical plus/minus 1 s envelopes for respective samples. Longest
timescale ($2^7$ x 512 ms) is 65 s.

\bigskip

\noindent {\bf Figure 2:} Similar to Fig. 1, for simulated dilated (dotted)
bursts and nondilated (long dash) bursts. Dilation factor $=$ 2.25, 100
bursts per group.

\bigskip

\noindent {\bf Figure 3:} Average wavelet-thresholded profiles of BATSE
bursts in three brightness groups, with highest peak for each burst shifted
into temporal alignment.  Dimmest (solid, outer profile), dim (dotted), and
bright groups (solid, inner profile).

\bigskip

\noindent {\bf Figure 4:} Similar to Fig. 3.  Average aligned,
wavelet-thresholded profiles for 100 simulation sets each of dilated
(dotted) and nondilated (long dash) bursts, 50 bursts per set.  Dilation
factor $=$ 2.25.  Envelopes (solid lines) contain 68\% of the profiles per
group.

\vfill\eject

\end